\documentstyle[12pt]{article}
\begin{document}
\vskip-1.0cm
\rightline{OITS-568}
\vskip.5cm

\centerline{\bf SELF-ORGANIZED CRITICALITY}
\centerline{\bf IN QUARK-HADRON PHASE TRANSITION}
\vskip.5cm
\centerline{Rudolph C. Hwa$^a$ and Jicai Pan$^b$}
\vskip.3cm
\centerline{$^a$Institute of Theoretical Science and Department of Physics}
\centerline{University of Oregon, Eugene, Oregon 97403}
\vskip.3cm
\centerline{$^b$Department of Physics, McGill University}
\centerline{Montreal, Quebec, Canada  H3A 2T8}

\begin{abstract}   The problem of clusters growth in
quark-hadron phase transition in heavy-ion collision is investigated by
cellular
automata.  The system is found to exhibit self-organized criticality with
the
distribution of cluster sizes having universal scaling behavior.
\end{abstract}

\vskip.5cm
This paper describes an unconventional study of an unconventional
signature of quark-gluon plasma.  It concerns the properties of the hadrons
produced after phase transition.  It is possible that the proposed
signature
might be suppressed by factors not yet considered, but if it survives,
there are
no competing processes that can cause ambiguity in the interpretation of
what
is observed.  For the moment the issue is what to expect under the best of
circumstances.

The problem is about the growth of hadronic clusters in the mixed phase,
which we take to be in an annular ring in a 2D section at midrapidity of
the
expanding cylinder after a head-on heavy-ion collision.  Hadrons, after
nucleation, move radially outward in the annular ring and can encounter
one another, resulting possibly in coalescence and in the formation of
hadronic clusters, if it is energetically favorable to reduce the surface
area.
Growth without collisions is also possible.  The clusters need not be
spherical.
Depending on factors that we know very little about dendritic or porous
structures cannot be ruled out.  Since the clusters in the quark
environment
behave much like massive colloids suspended in a fluid, they carry out
Brownian motion, in the course of which the clusters collide and grow in
size.  Because nucleation can take place anywhere in the mixed region at
any
time during the period when the mixed phase exists, the clusters emitted at
the boundary of the mixed region can have various sizes $S$, depending on
how far the nucleation position is from the boundary.  A determination of
the distribution $P(S)$ of the cluster size would therefore reveal the
various
aspects of the dynamical process of hadronization mentioned above.

There is no reliable way to treat the cluster growth problem analytically.
Our approach is to use cellular automata based on simple rules to simulate
the dynamical evolution of the problem.  Instead of making drastic
approximations necessary to put the problem on a continuum basis amenable
to analytical solution, we confront the complications by finding a set of
rules that capture the essence of the dynamics.  This has been done both in
1D \cite{rch4} and in 2D \cite{rch5}.  We refer the reader to those
references
for the details of the precise rules.

In 2D a wedge of the annular ring containing the mixed region is mapped to
a
square lattice of $L$ sites in each dimension initially with periodic
boundary
condition on the upper and lower sides. The nucleation size $S_0$ is
specified
in units of site separation.  The complication of confinement at the local
level
is summarized by one parameter $p$, the nucleation probability.  At each
time step an unoccupied site (quark) has probability $p$ of becoming
occupied (hadron).  The occupied sites all take random walks superimposed
on an average drift to the right boundary.  Clusters of occupied sites
become
bonded when they meet and become larger clusters.  This process is
repeated again and again.  A cluster of size $S$ reaching the right
boundary is
taken out of the lattice, and $S$ vertical units of the boundary are moved
to
the left by one unit.  When the moving right boundary reaches the fixed
left
boundary, the phase transition is over.

Repeated simulation using the automaton thus constructed results in the
distribution $P(S)$ that has a power-law behavior
		\begin{eqnarray}
P(S) \propto S^{-\gamma} \quad
\label{1}
  	\end{eqnarray}
as shown in Fig.\,1(a), for $L = 16$, $S_0 = 1$, and various values of $p$.

The same behavior persists when $L$ is changed to 32 [Fig.\,1(b)], and when
$S_0$ is changed to 2 [Fig.\,1(c)].  Thus the exponent $\gamma$ is
independent of $L$ and $S_0$, and is rather insensitive to $p$, since for
$p$
ranging over an order of magnitude from 0.05 to 0.5, $\gamma$ varies only
within 10\% of the average value $\gamma \simeq 1.9$.  This result on the
cluster-size distribution is therefore both scaling and universal.

Thermal equilibrium has not been assumed. The process is non-equilibrium
and dissipative.  The system evolves on its own to a critical point
characterized by large fluctuations of all sizes.  Most significantly, no
parameters have been tuned to put the system at the critical point.  Such
behaviors belong to the class of systems that exhibit what has come to be
referred to as self-organized criticality (SOC) \cite{pb}.  Examples of SOC
are
the sandpile, forest fire, earthquake and other large scale problems.  In
the
sandpile problem the distribution of avalanche sizes satisfies a scaling
law,
just like (\ref{1}), and is universal \cite{pb}.  The use of cellular
automata
and the finding of (\ref{1}) for quark-hadron phase transition put the
physics of heavy-ion collisions for the first time in touch with modern
trends in statistical physics.

There are two ways in which the rules in \cite{rch5} can be improved.  So
far the clusters formed are rigid and have irregular shapes, whereas the
possibility to deform seems reasonable.  Furthermore, the breakup of a
cluster should be allowed, as is done in 1D \cite{rch4}, but too
complicated to
implement in 2D \cite{rch5}.  Both of these improved features have been
introduced recently in a new cellular automaton, in which a cluster is
treated as a vertical stack located at one site, having an effective
radius of extension
   \begin{equation}
R = \alpha \sqrt{S/\pi}
\label{2}
\end{equation}
where $\alpha$ is a number that parametrizes the shape of the cluster
\cite{rch6}.  A cluster with $\alpha >1$ would be dendritic, while $\alpha
=1$ corresponds to a circle.  Breakup $(S \rightarrow S_1 + S_2)$ is
allowed
to take place with the probability
\begin{equation}
B(S, S_1, S_2) = \beta \left( \sqrt{S_1/S} + \sqrt{S_2/S} \right)^{-1}
\label{3}
\end{equation}
where $\beta$ is another parameter.  We have found that when $p > 0.05$,
$P(S)$ is insensitive to the value of $\alpha$ for $1.0 < \alpha < 1.4$.
The
dependence on $\beta$ is shown in Fig.\,1(d).  Evidently, not only is the
scaling law unchanged, there is virtually no dependence on the breakup
parameter $\beta$, when $p$ is not too small.  It means that with
sufficient
nucleation and growth the broken pieces can readily recombine to
form larger clusters.

The clusters that we have considered are
produced at the edge of the mixed-phase region still at the transition
temperature.  What changes can take place as they cool to $T = 0$ on their
way to the detector is a different issue that cannot be addressed
adequately here.  If there is no hadron gas surrounding the plasma, there
would be no collisions among the clusters, which then simply travel in
free streaming toward the detector.  To separate the various clusters
experimentally, we suggest that the $p_T$ space be partitioned into
many bins so that in each $p_T$ bin one analyzes the particle density in
small
bins in the $\eta$-$\phi$ space in search for clustering.  In that way the
various clusters produced in an event do not overlap and defeat
identification.
Of course, it is necessary to make event-by-event analysis, lest all
signatures
would be smeared out in an inclusive distribution.

In summary, our theoretical study has led to the result
that a universal scaling behavior
of cluster production is a likely feature at the end of the quark-hadron
phase transition.  The important condition under which this result follows
is
that there is an extended region of mixed phase that persists for a long
time.
Apart from that, there is approximate
independence on all other experimental conditions. The existence of large
scale
fluctuations in extended systems is just the feature that was discovered in
the
study of self-organized criticality.  Our finding therefore connects
heavy-ion
collisions to a very new area of research in statistical physics.
Although only
theoretical at this stage, the result should provide sufficient motivation
to
search experimentally for the clustering of hadrons, the discovery of which
in
the data would undoubtedly be a stimulating advance in the creation of
quark-gluon plasma.

\vskip.7cm
\centerline{\bf Acknowledgment}
\medskip
This work was supported in part by the U.S. Department of Energy under
Grant No. DE-FG06-91ER40637, and by the Natural Science and Engineering
Council of Canada and by the FCAR fund of the Quebec Government.

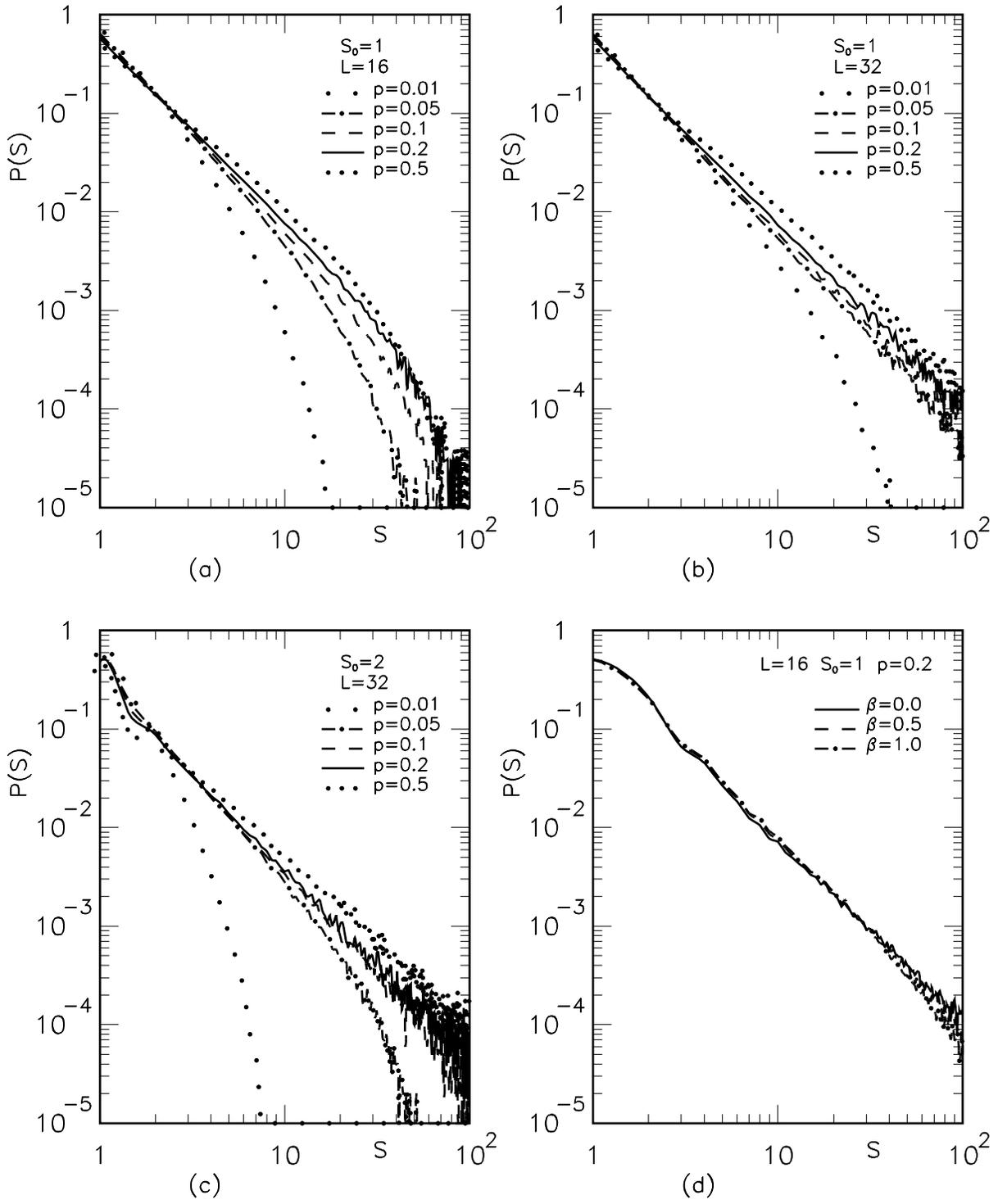
\end{document}